\documentclass[aps,prl,reprint,twocolumn,amsmath,amssymb,citeautoscript]{revtex4-1}
\usepackage{graphicx}% Include figure files
\usepackage{dcolumn}% Align table columns on decimal point
\usepackage{bm}% bold math
\usepackage{latexsym,epsfig}
\usepackage{graphicx}
\usepackage{verbatim}
\usepackage{comment}
\usepackage{amsmath}
\usepackage{amssymb}
\usepackage{physics}
\usepackage{stmaryrd}
\usepackage{color}
\usepackage{epstopdf}
\usepackage{grffile}
\usepackage{lipsum}
\usepackage{enumitem}
\usepackage[normalem]{ulem}
\DeclareGraphicsExtensions{.eps}

\newcommand{\beq}{\begin{equation}}
\newcommand{\eeq}{\end{equation}}
\newcommand{\bea}{\begin{eqnarray}}
\newcommand{\eea}{\end{eqnarray}}
\newcommand{\ben}{\begin{eqnarray*}}
\newcommand{\een}{\end{eqnarray*}}
\newcommand{\bfig}{\begin{figure}}
\newcommand{\efig}{\end{figure}}

\usepackage{hyperref}
\hypersetup{
    colorlinks=true,      
    urlcolor=blue,
    citecolor=blue,
    linkcolor=blue
}

\begin{document}
\title{Re-entrant topological phase transition in a non-Hermitian quasiperiodic lattice} %\padhan{quasicrystal instead of quasiperiodic lattice?}}
\author{Ashirbad Padhan$^1$, Soumya Ranjan Padhi$^{2, 3}$ and Tapan Mishra$^{2,3,*}$}
 \affiliation{$^{1}$Department of Physics, Indian Institute of Technology, Guwahati, Assam - 781039, India}

\affiliation{$^2$ School of Physical Sciences, National Institute of Science Education and Research, Jatni 752050, India}

\affiliation{$^3$ Homi Bhabha National Institute, Training School Complex, Anushaktinagar, Mumbai 400094, India}

 \email{mishratapan@niser.ac.in}
\date{\today}

\date{\today}

\begin{abstract}

We predict a re-entrant topological transition in a one dimensional non-Hermitian quasiperiodic lattice. By considering a non-Hermitian generalized Aubry-Andr\'e-Harper (AAH) model with quasiperiodic potential, we show that the system first undergoes a transition from the delocalized phase to the localized phase and then to the delocalized phase as a function of the hermiticity breaking parameter. This re-entrant delocalization-localization-delocalization transition in turn results in a re-entrant topological transition identified by associating the phases with spectral winding numbers.
Moreover, we find that these two transitions occur through intermediate phases hosting both extended and localized states having real and imaginary energies, respectively. We find that these phases also possess non-trivial winding numbers which are different from that of the localized phase.

\end{abstract}

%% APS is now using PhySH!!
%\pacs{67.80.K-, 75.10.Jm, 03.75.Lm}

% 75.40.Gb	Diffusion spin, Magnetic properties (spin waves, dynamic critical point effects)
% 67.85.-d	Ultracold gases, trapped gases, 
% 71.27.+a	Heavy-fermion solids (electron states), Strongly correlated electron systems

% 67.85.Fg	Condensates (spinor condensates)
% 64.70.Tg	Quantum phase transitions
% 67.85.Bc	Condensates (static properties of)

% 67.80.K-	Supersolids quantum solids
% 75.10.Jm	quantum spin frustration, Quantized spin models, Heisenberg model, Hubbard model, magnetic ordering (quantized spin model
% 03.75.Lm	Tunneling, Josephson effect, Bose-Einstein condensates in periodic potentials, solitons, vortices, and topological excitations
% 67.85.Jk	Bose-Einstein condensates
% 75.50.Ee	Antiferromagnetics

\maketitle
%\section{Introduction} 
{\em Introduction.-} 
%The simplest such example is the one dimensional non-Hermitian Aubry-Andre model which is a  In particular, the non-Hermitian quasiperiodic lattice in one dimension  
Quasiperiodic lattices which are intermediate to periodic and random lattices have enriched our understanding on the localization transition~\cite{paredesreview}. Especially in one dimension, the quasiperiodic lattice systems exhibit well defined localization transition as opposed to their random counterparts where the localization of states occurs for an infinitesimal strength of disorder. One of the simplest quasiperiodic lattice models is the paradigmatic Aubry-Andr\'e-Harper (AAH) model which exhibits a sharp delocalization-localization (DL) transition at a critical quasiperiodic potential strength~\cite{aubry1980analyticity,P_G_Harper_1955}. Apart from the point of view of localization transition, AAH model can also be appropriately connected to the quantum Hall systems exhibiting topological character such as well defined bulk topological invariant and conducting edge modes. Further generalizations of the AAH model have resulted in a great deal of novel scenarios in the context of localization transition in recent years~\cite{subroto_review}. One of the important manifestations of such generalization is the  localization transition through an intermediate phase with coexisting  extended and localized states which are separated by an energy dependent mobility edge~\cite{NN_TB_sarma2015,Tri_pot_ladder_AA,ME_bich_sarma_2017,expt2,ME_Incomm_Opt_sarma2010}. 
%On the other hand Owing to their simpler form, various studies have been performed based on the AA model and its variants both theoretically~\cite{Loc_AA_non-nerest,ME_IP_sarma2020,ME_Incomm_Opt_sarma2010,NN_TB_sarma2015,add_many_papers} and experimentally~\cite{expt1,expt2,expt_Int_ME_GAA} including in the context of the many-body localization transition~\cite{mbl_quasi_periodic}. 

On the other hand, non-Hermitian quasiperiodic lattices offer a much richer paradigm for the study of localization transition as compared to their Hermitian counterpart. Numerous studies have been performed on the non-Hermitian AAH model by introducing the non-Hermiticity through the onsite potential or through the non-reciprocal hopping. Incorporating such terms in the AAH model results in a sharp DL transition, associated $\mathcal{P}\mathcal{T}$ symmetry breaking, butterfly spectra, non-Hermitian mobility edges, appearance of topological edge states etc~\cite{Top_phases_NHAAH,Jazaeri_2001_loc_incom,LOC_TOP_NHQP_Lattice,BD_SDs_WN_NHAP1d_model,Int_SEs_AL_QP,Pha_tra_NHAAH,NHSE_WN_DNHS,NHAAH_GL_2017,WN_GME_NHS,Top_phases_NHAAH,Robustness_eigenspectra_shuchen,Slowly_vraying_potential,Symm_top_NHP,Edge_ST_top_inv_NHS,Butterfly_Yuce_2014,MBL_non_her_QPs}. A recent study on a AAH model with complex phase has shown that the localized phase where the entire energy spectrum is complex can be associated with a spectral winding number~\cite{longhi_PRL} differentiating it from the delocalized phase where the winding number is zero. This results in a topological transition as a function of the Hermiticity breaking parameter in the quasiperiodic AAH model. Similar to the case of the Hermitian generalized AAH model, the non-Hermitian generalized AAH (nHGAAH) models also exhibit an intermediate region across the DL transition hosting a mobility edge which also separates the states having real and complex energies~\cite{decaying_hopp_NHQC,NHAAM_power_law_hopp,FLOQ_TOP_LOC_MEs_1d_NHQCs,Loc_PT_sym_breaking,NHQW_zhou_2021non,LOC_TOP_NHAAH_p-wave,Dri_mult_PT_re-entrant_Floq_QCs,Exp_Top_pha_MEs_QCs,Breakdown_RC_LD_NHQCs}. Following the prescription provided in~\cite{GAA_exp}, this intermediate region can be associated with a well defined winding number which makes them topologically non-trivial. This interesting manifestation of the delocalization-localization transition exhibiting real-complex as well as trivial-topological transition has recently attracted a great deal of attention to understand the DL transitions in the variants of the nHAAH models~\cite{Dynamic_PT_symm,Dyn_loc_NHQCs,MI_NHAAH,GAA_exp,short_range_GAA,mosaic_PT_shuchen}. Due to the possibility of accessing such systems in artificial systems such as photonic lattices~\cite{GAA_exp}
and electrical circuits~\cite{WN_GME_NHS} and the recent experimental observation of such DL transition in a driven systems~\cite{Triple_phase_transition} have paved the path for further exploration in the field.

While in the previous studies based on the AAH model with complex quasiperiodic disorder exhibit only the DL transitions, in this paper we show that in the case of a generalized AAH model with complex quasiperiodic potential the system returns to the delocalized phase after undergoing a DL transition as a function of the hermiticity breaking parameter. Due to the complex spectrum, we identify the localized phase as topologically non-trivial where a spectral winding number can be defined which vanishes in the delocalized phase due to the complete real spectrum. This results in a re-entrant topological transition in the systems. However, contrary to the case of the non-Hermitian AAH model considered in Ref.~\cite{longhi_PRL}, in this case the DL and LD transitions occur through the intermediate or mixed regions. We also obtain that the system is topological when in these mixed regions due to the presence of complex eigenvalues in the spectrum.  In the following we discuss these findings in detail. 

{\em Model.-} 
The non-Hermitian generalized AAH model is defined as 
\begin{align}
    H = -J \sum_{n=1}^{L} \big{(}c_n^\dagger c_{n+1} + H.c.\big{)} \nonumber\\
    +  \lambda \sum_{n=1}^{L} \frac{\cos(2\pi\beta n + \phi)}{1 - \alpha \cos(2\pi\beta n + \phi)} c_n^\dagger c_n,
\label{eq:ham}
\end{align} 
where $c_n^\dagger~(c_n)$ is the creation (annihilation) operator of spinless fermions at the $n$th lattice site. $J$ is the nearest-neighbor hopping amplitude and $\lambda$ represents the strength of the  quasiperiodic potential. Here $\beta = (\sqrt{5} - 1)/2$ - an irrational number known as the inverse golden ratio and $\phi$ is the phase. The non-Hermiticity in the system is introduced by defining a complex phase $\phi=\theta+i h$. Note that the nHGAAH respects $\mathcal{P}\mathcal{T}$-symmetry if we choose the real part of the phase to be zero. Therefore, we consider $\theta=0$ throughout the paper, unless otherwise explicitly mentioned. 

{\em Results.-}
For $\alpha=0$ and $h=0$, the model reduces to the Hermitian AAH model that exhibits a delocalization-localization transition of the entire spectrum at $\lambda=2J$ due to the self-duality of the model. However, when $h$ is finite, the model is non-Hermitian and has been studied in Ref.~\cite{longhi_PRL} predicting a delocalization-localization transition of the entire spectrum  at a critical $h=\ln(2J/\lambda)$ that also coincides with a $\mathcal{P}\mathcal{T}$ symmetry breaking phase transition indicated by a real-complex transition of the entire energies. Furthermore, it has been shown that this transition is topological in nature  characterized by a spectral winding number which is zero (one) in the $\mathcal{P}\mathcal{T}$ unbroken (broken) phase. Altogether, one gets a triple phase transition at $h=\ln(2J/\lambda)$ when $\alpha=0$. However, when $\alpha$ becomes finite, a completely different scenario appears. In the following we show that the system first undergoes a triple phase transition and then at a later stage the system returns to its original state undergoing another triple phase transition as function of $h$. These findings are obtained by numerically solving the model shown in Eq.~\ref{eq:ham} using the exact diagonalization method under periodic boundary conditions (PBCs) with systems of size up to $L=6765$. We set $J=1$ as the energy scale and fix the strength of the quasiperiodic potential $\lambda=1$. 
\begin{figure}[t]
\centering
\includegraphics[width=1\columnwidth]{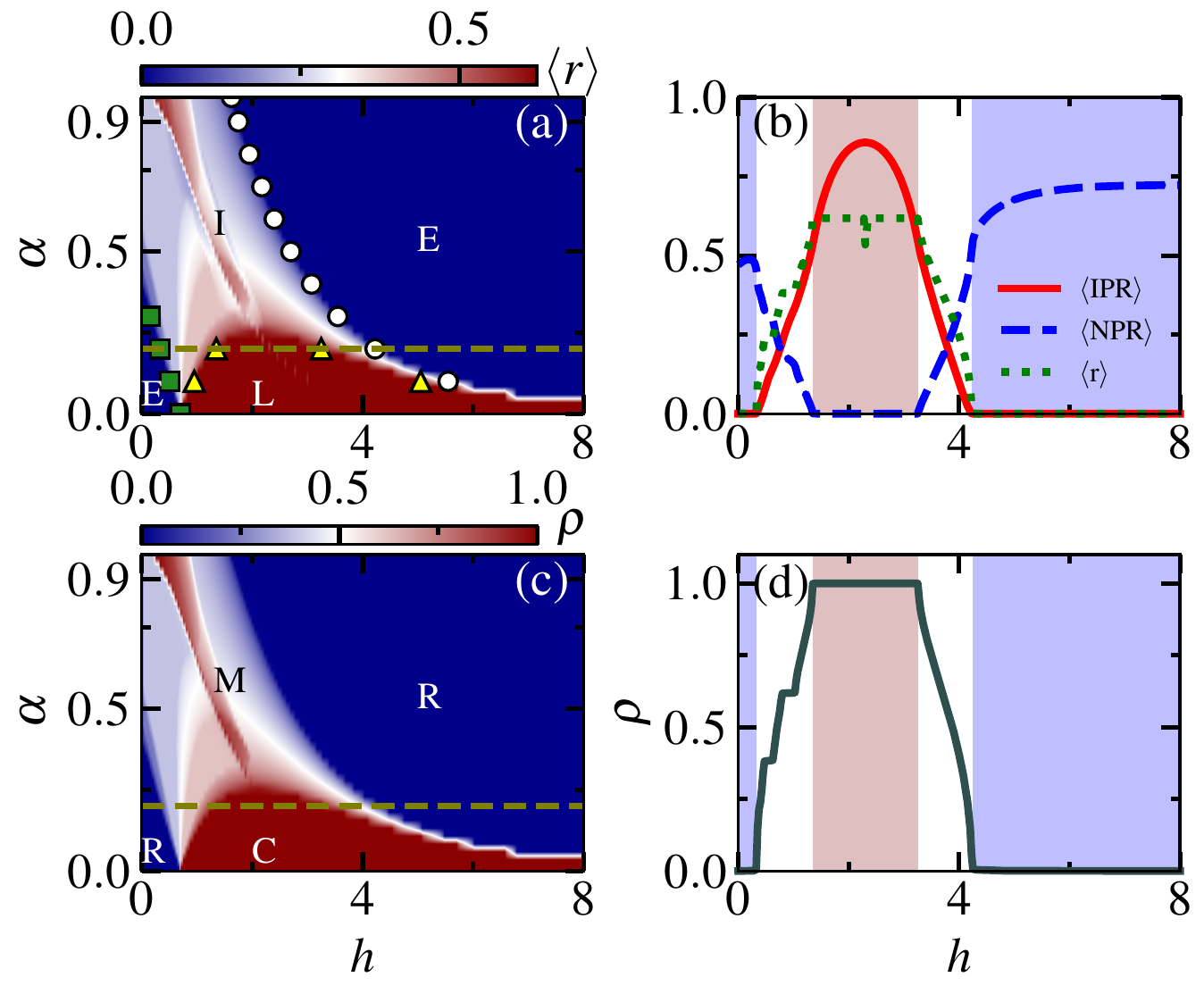}

\caption{(a) Phase diagram in the $\alpha$-h plane, obtained using the $\langle \rm IPR \rangle$ and $\langle \rm NPR \rangle$ values indicating the extended (E), intermediate (I) and localized (L) phases. The color code indicates the values of $\langle r \rangle$ superimposed in the figure. (b) $\langle \rm IPR \rangle$ (solid red line), $\langle \rm NPR \rangle$ (dashed blue line) and $\langle \rm r \rangle$ (dotted green line) are plotted as a function of $h$ for $\alpha=0.2$, indicating a re-entrant localization transition. (c) Phase diagram $\alpha-h$ plane obtained using the density of states $\rho$ which distinguishes the $\mathcal{P}\mathcal{T}$ unbroken (R), mixed (C) and broken (R) phases. (d) $\rho$ as a function of $h$ for $\alpha=0.2$ indicating a real-complex-real transition. The shaded blue (red) areas in (b) and (d) denote the extended (localized) and $\mathcal{P}\mathcal{T}$ unbroken (broken) phases, respectively. Here the system size is considered as $L=6765$.}
\label{fig:fig1}
\end{figure}

In the following, we discuss our findings in detail. First we will focus on the delocalization-localization transition. Next we will investigate the transition related to the $\mathcal{P}\mathcal{T}$ symmetry breaking and then explore the spectral topological character associated to these transitions. 

{\em Delocalization-localization transition,-}
We begin our discussion by identifying the delocalized and localized regions in the $h-\alpha$ plane as shown in Fig.~\ref{fig:fig1}(a). The regions below the boundary with green squares and above white circles correspond to extended phase (E) and the region below the yellow  triangles corresponds to the localized phase (L). The white central region enclosed by these three lines is the intermediate region (I) where both extended and localized states coexist. These boundaries are obtained from the inverse participation ratio (IPR), given by ${\rm{IPR}}_{n}=\sum_{j=1}^{L} |\psi^{j}_{n}|^{4}$  and the corresponding normalized participation ratio (NPR) given by ${\rm{NPR}}_{n}=1/(L\times {\rm{IPR}}_{n})$~\cite{short_range_GAA, GAA_exp} where $\psi^{j}_{n}$ is the n$^{th}$ eigenstate of the Hamiltonian shown in Eq.~\ref{eq:ham}. The ${\rm{IPR}}$ (${\rm{NPR}}$) takes vanishing (finite) and finite (vanishing) value for an extended and a localized state, respectively for a finite system. However, to obtain the insight about the entire spectrum, we utilize the average values of ${\rm{IPR}}$  and ${\rm{NPR}}$ taken over all the states. 
In Fig.~\ref{fig:fig1}(b), we plot $\langle {\rm{IPR}}\rangle$ (red solid line) and $\langle {\rm{NPR}}\rangle$ (blue dashed line) as a function of $h$ for an exemplary value of $\alpha=0.2$. Here $\langle \cdot \rangle$ stands for the average over all eigenstates. The values of $\langle {\rm{IPR}}\rangle=0$  and $\langle {\rm{NPR}}\rangle \neq 0$ in the regions $h < 0.325$ and $h > 4.25$ (light blue regions in Fig.~\ref{fig:fig1}(b)) indicate that all the states in the system are extended. In the range $1.35 < h < 3.25$ (light red region), the states are localized which is indicated from the values $\langle {\rm{IPR}}\rangle \neq 0$  and $\langle {\rm{NPR}}\rangle = 0$. However, there exist two intermediate regions on either sides of the localized region where both $\langle {\rm{IPR}}\rangle$ and $\langle {\rm{NPR}}\rangle$ remain finite (white regions between $0.325 < h < 1.35$ and $3.25 < h < 4.25$ in Fig.~\ref{fig:fig1}(b). The boundaries in the phase diagram shown in Fig.~\ref{fig:fig1}(a) are obtained by plotting the average participation ratios at different values of $\alpha$.

It can be noticed from the phase diagram that when $\alpha=0$, a sharp delocalization to localization transition occurs at $h=\ln(2J/\lambda)$ - a feature already predicted in Ref.~\cite{longhi_PRL}. However, as $\alpha$ increases, the system undergoes a delocalization-localization-delocalization transition for a range of $\alpha$ indicating a re-entrant transition. Unlike the transition at $\alpha=0$, these transitions occur through the intermediate regions and are not sharp. Note that for higher values of $\alpha$ i.e. $\alpha \gtrsim 0.25$, the localization transition does not occur and the re-entrant transition occurs through the intermediate regions only. Further increase in the value of $\alpha$ results in a direct transition from the intermediate to extended phase. In these limits of $\alpha$, the system remains in the intermediate phase for a range of $h$ starting from $h=0$. 

To further quantify this behaviour of delocalized-intermediate-localized-intermediate-delocalized transition, we compute the adjacent gap ratios (AGRs) defined by $r_n=\frac{\rm{min}(\epsilon_n, \epsilon_{n+1})}{\rm{max}(\epsilon_n, \epsilon_{n+1})}$ where $\epsilon_n=\Re(E_n) - \Re(E_{n-1})$.  Note that here the eigenvalues $E_n$ are sorted in ascending order according to their real part only~\cite{Dimer_multi_loc,Multi_reentrant}. In Fig.~\ref{fig:fig1}(b), we plot the average AGR i.e. $\langle r\rangle=\sum_nr_n/L$ (green dashed line) as a function of $h$ for $\alpha=0.2$. As expected, $\langle r\rangle$ vanishes in the extended phase and attains its maximum value in the localized phase but takes an intermediate value in the intermediate phases. To identify the phases from the behaviour of $\langle r\rangle$ we plot $\langle r\rangle$ as a function of $\alpha$ and $h$ in Fig.~\ref{fig:fig1}(a). The delocalized and localized regions can be clearly identified by the blue and red  region where $\langle r\rangle$ is zero and $\langle r\rangle$ finite respectively. This also matches well with the boundaries obtained from the average participation ratios [symbols in Fig.~\ref{fig:fig1}(a)]. We further notice that $\langle r\rangle$ attains a value $\sim0.5$ (sharp dip) for a particular value of $h_c = \ln|\frac{1+\sqrt{1 - \alpha^2}}{\alpha}|\sim2.292$ inside the localized region. Such a peculiar behavior at this point will be discussed later. 

The above analysis clearly shows that increase in $h$ turns all the extended states localized and then extended again leading to a re-entrant delocalization. These transitions occur through two intermediate regions. In the following we will analyse the $\mathcal{P}\mathcal{T}$ breaking transition associated to these re-entrant transition points.

\begin{figure}[t]
\centering
\includegraphics[width=1\columnwidth]{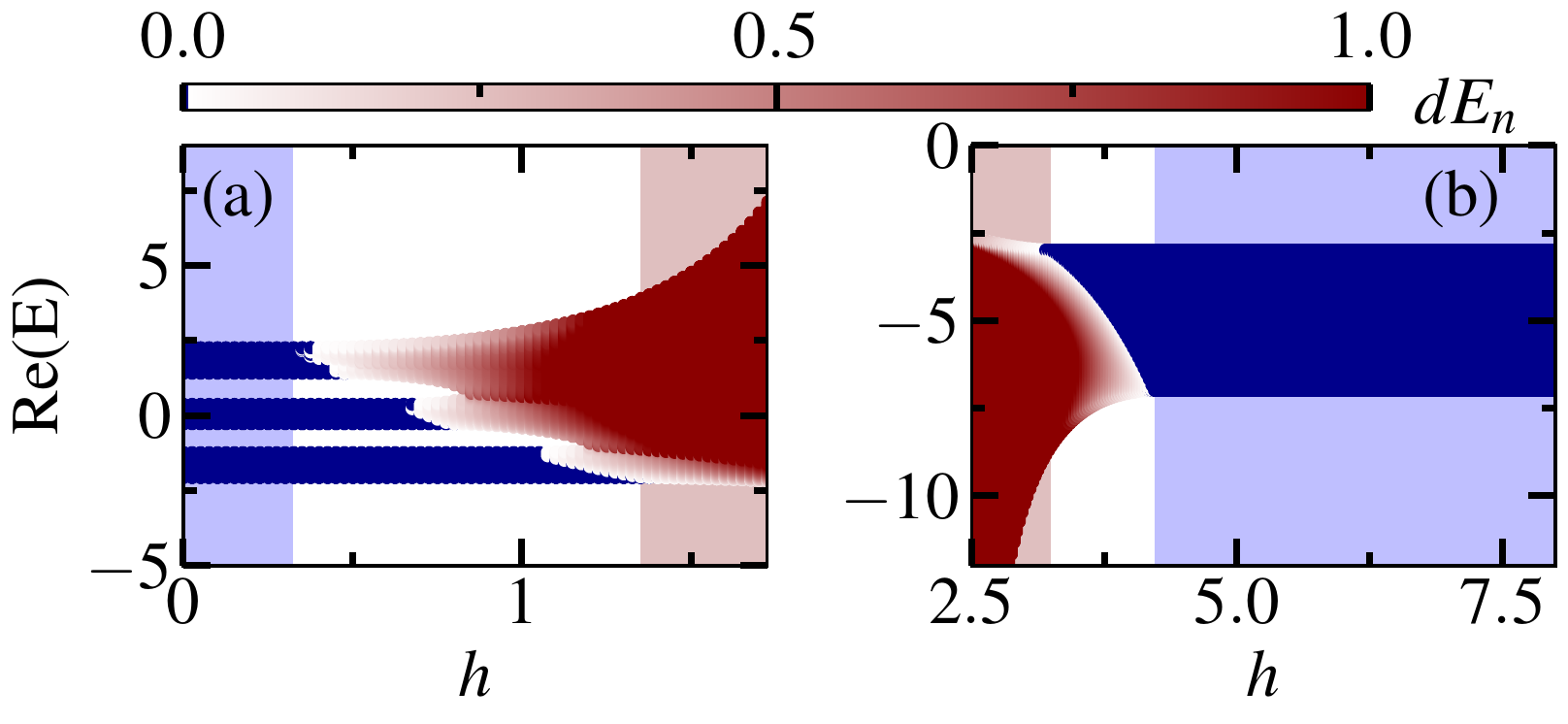}
\caption{(a) $dE_n(0)$ and (b) $dE_n(8)$ are plotted as a function of real eigenvalues and $h$ for $\alpha=0.2$ and $L=2584$, showing the robustness of the spectrum in the extended phases. The color shows the values of $dE_n$ which is rescaled for clarity. }
\label{fig:fig3}
\end{figure}

{\em Real-complex transition.-}
As mentioned earlier, the model shown in Eq.~(\ref{eq:ham}) is $\mathcal{P}\mathcal{T}$ symmetric\cite{bender_prl_1998, nature_PT_review_2018, PT_photonics_review} when $h=0$. In the limit $\alpha=0$, the  Eq.~\ref{eq:ham} becomes the AAH model and exhibits a delocalization-localization  transition which coincides with a real-complex transition in the spectrum as a function of $h$~\cite{longhi_PRL}. This real-complex transition is associated to the $\mathcal{P}\mathcal{T}$ symmetry breaking in the system~\cite{MI_NHAAH, GAA_exp, short_range_GAA, mosaic_PT_shuchen}. To investigate the spectrum in the current scenario, i.e when $\alpha$ is finite, we analyse the behaviour of the density of states $\rho=N/L$, where $N$ counts the number of states having complex eigenvalues in the spectrum and $L$ is the system size. According to the definition, in the thermodynamic limit, $\rho$ attains the value $0~(1)$ when none (all) of the eigenenergies are complex and the corresponding phase is $\mathcal{P}\mathcal{T}$ unbroken (broken). To this end, we first plot $\rho$ as a function of $\alpha$ and $h$ in Fig.~\ref{fig:fig1}(c) which clearly depicts the regions of real energies (blue region) and complex energies (red region) where $\rho$ becomes exactly zero and one respectively. There also exists a region where the value of $\rho$ is in between zero and one which indicates the presence of both real and imaginary eigenenergies in the spectrum and we call it the mixed region. Note that similar to the phase diagram shown in Fig.~\ref{fig:fig1}(a), we also see a re-entrant behaviour in Fig.~\ref{fig:fig1}(c). To understand this clearly, we plot $\rho$ as a function of $h$ in Fig.~\ref{fig:fig1}(d) for a cut through the phase diagram of Fig.~\ref{fig:fig1}(c) at $\alpha=0.2$ (depicted as the dashed line in Fig.~\ref{fig:fig1}(c)). From Fig.~\ref{fig:fig1}(d), we observe that initially $\rho=0$, i.e. all the energies are real in the spectrum up to $h=0.325$ (light blue region). As $h$ increases, the value of $\rho$ becomes finite and reaches its maximum, i.e. $\rho=1$ in the range $1.35 < h < 3.25$ (light red region). In this range of $h$, the entire spectrum is complex and the states are localized. Further increase in $h$ leads to a decrease in the value of $\rho$ which eventually becomes zero for $h>4.25$, after which the spectrum is real again (light blue region). This re-appearance of the entire real spectrum for large values of $h$ is an indication of a re-entrant real-complex-real transition in the spectrum which indicates a $\mathcal{P}\mathcal{T}$ unbroken-broken-unbroken phase transition. We also find that in between the two extreme values of $\rho$, there exist the mixed regions (marked by white regions in Fig.~\ref{fig:fig1}(d)) where $\rho$ takes values between $0$ and $1$. Note that the step wise increase of $\rho$ in the first mixed region is due to the gaps in the spectrum which is shown in Fig.~\ref{fig:fig3}. Comparing the behaviour of $\rho$ with the participation ratios shown in  Fig.~\ref{fig:fig1}(b), we obtain that the delocalized-intermediate-localized-intermediate-delocalized transitions of the eigenstates coincide with the real-mixed-complex-mixed-real transitions in the energies. Similar to the localization properties, the re-entrant real-complex transition occurs  for a small range of $\alpha$. After $\alpha \geq 0.25$, the spectrum is never entirely complex and the re-entrant transition is of real-mixed-real type for $0.25\leq\alpha\leq3.75$ and for $\alpha>3.75$ the transition is of mixed-real type as can be seen from Fig.~\ref{fig:fig1}(c).

This re-entrant real-complex-real transition can also be characterised by analysing the robustness of the real energy spectrum 
as the complex phase is varied~\cite{Robustness_eigenspectra_shuchen}. It is expected that the eigenvalues will remain constant as long as the states are delocalized or extended. This can be confirmed by computing the quantity $dE_n(h^{'}) = |E_n(h^{'})-E_n(h)|$ which defines the energy shift of the $n$th eigenvalue for a particular $h$ from that for a fixed $h^{'}$. In Fig.~\ref{fig:fig3}(a) and (b) we plot $dE_n(h^{'}=0)$ and $dE_n(h^{'}=8)$, respectively, as a function of real eigenvalues (Re$(E)$) and $h$ for $\alpha=0.2$. As expected, when the system is in the delocalized phase, i.e., up to $h<0.325$ and $h>4.25$, we obtain $dE_n(h^{'})=0$ (the dark blue regions) for all the eigenstates indicating the robustness of the real spectrum. However, in the range $0.325 < h < 4.25$, the values of $dE_n(h^{'})$ become positive for some of the states, which is a signature of the appearance of the localized states in the system resulting in a mixed spectrum.

The above analysis shows that the nHGAAH model exhibits a delocalized-localized-delocalized type re-entrant transition of eigenstates that simulataneously occurs with  a real-complex-real transition of the eigenspectrum which is associated to the $\mathcal{P}\mathcal{T}$ symmetry unbroken-broken-unbroken phase transition. Note that this re-entrant transition is not observed for $\alpha=0$ as can be seen from Fig.~\ref{fig:fig1}(a) and (c) and also predicted in Ref.~\cite{longhi_PRL}.

\begin{figure}[t]
\centering
\includegraphics[width=1\columnwidth]{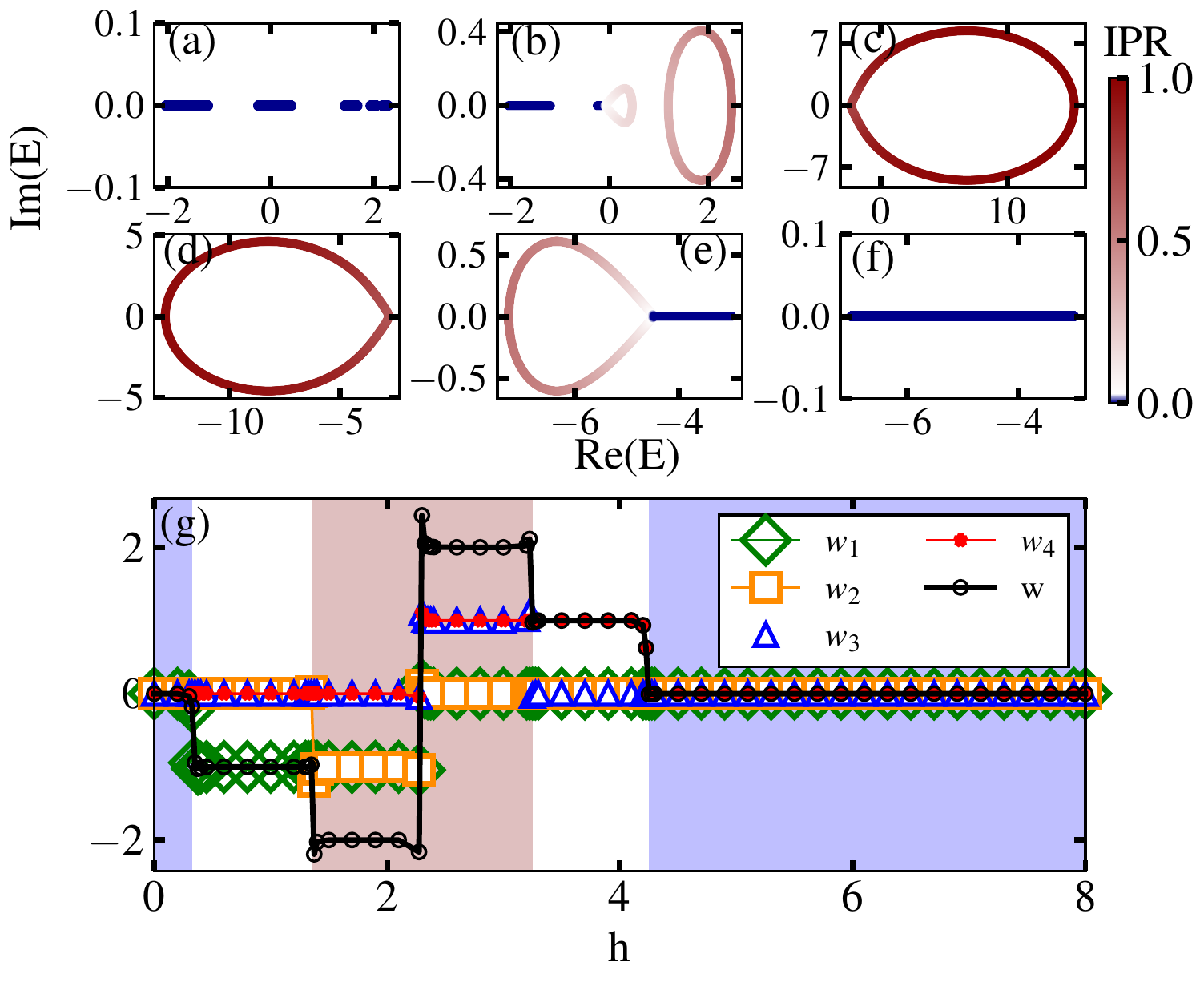}
\caption{(a-f) IPR values plotted as a function of real and imaginary eigenvalues corresponding to $h=0.2, 0.75,2.0, 2.8, 3.75$ and $5.0$, respectively for $\alpha=0.2$. (g) shows the variation of spectral winding numbers $w_1$ (green diamonds), $w_2$ (orange boxes), $w_3$ (blue triangles) and $w_4$ (red filled circles) and total winding number $w$ (black empty circles) as a function of $h$ for $\alpha=0.2$. The shaded blue (red) area in (g) denotes the trivial (nontrivial) phase. Here the system size is considered to be $L=2584$ in (a-f) and $L=233$ in (g).}

\label{fig:fig2}
\end{figure}

{\em Topological transition.-}
In this part of the paper, we will identify the different phases with respect to their topological nature. As mentioned before, the localized phase in the nHAAH model is topological which possess a non-trivial spectral winding number that is derived  from the winding of the complex spectral trajectory around certain base energy~\cite{GAA_exp,Top_phases_NHS}. The winding number is defined as ~\cite{longhi_PRL}
\begin{align}
    w=\lim_{L\to\infty} \frac{1}{2\pi i}  \int_{0}^{2\pi}d\theta\partial_{\theta}\log\big[\det\{H(\theta/L)-\varepsilon\}\big],
\label{eq:wind}
\end{align} 
where, $\varepsilon$ are the base energies. A winding number in this case is defined as the number of times the spectrum of $H$ winds the base energy when the real $\theta$ varies from $0$ to $2\pi$. In the case of the nHAAH model, there is a direct real-complex transition in the spectrum as a function of $h$ i.e. all the energies in the spectrum become complex. This relaxes the choice of the base energy which can be safely taken to be zero~\cite{longhi_PRL}. However, in the presence of the mobility edge between the extended and the localized regions, the base energy can not be arbitrary due to the presence of the mixed states in the spectrum. In practice, two base energies are considered which correspond to the real energy eigenvalues that defines the beginning and the end of the intermediate or mixed  region or the minimum and maximum energies on the mobility edge~\cite{GAA_exp, NHQW_zhou_2021non}. However, since our system exhibits two intermediate phases and a localized phase hosting complex spectra and resulting in four transition points in total, we need to define four winding numbers. Here we compute the winding numbers while crossing through all the phases at $\alpha=0.2$. First we plot the real and imaginary energies in Fig.~\ref{fig:fig2} (a-f) for different values of $h$ (i.e. $h=0.2, 0.75, 2.0, 2.8, 3.75$ and $5.0$). This shows that when in the delocalized phase, all the energies are real (see Fig.~\ref{fig:fig2}(a) and (f)). However, in the localized and the intermediate phases, complex eigenvalues appear which form loops as shown in Fig.~\ref{fig:fig2}(b-e) for $h=0.75, 2.0, 2.8$ and $3.75$. We identify the winding numbers corresponding to these loops as follows. We first identify the base energies at the beginning and at the end points of the mobility edge from the energy spectrum. For $\alpha=0.2$, the base energies are $E_1\approx 2.274$ and $E_2\approx -2.042$ at the critical points $h_1$ and $h_2$ respectively. Similarly, we fix two other base energies across the second intermediate region i.e. $E_3\approx -3.0$ and $E_4\approx -7.0$ at $h_3$ and $h_4$ respectively. Accordingly, we obtain four winding numbers such as $w_1$ (green diamonds), $w_2$ (orange boxes), $w_3$ (blue triangles) and $w_4$ (red filled circles) using Eq.~\ref{eq:wind}  which are plotted as a function of $h$ in Fig.~\ref{fig:fig2} (g). We also show the total winding number $w=w_1+w_2+w_3+w_4$ (black circles) to clearly distinguish the different phases.
As observed, all the winding numbers vanish in the delocalized phase due to the non-existence of complex eigenvalues. In the intermediate phase, one of the winding numbers is finite, (e.g. for $0.325<h<1.35$, $w_1=-1$) and in the localized phase, two of them are finite (e.g. for $1.35<h<h_c$, $w_1=w_2=-1$). However, a counter-intutive situation arises at $h_c$ where the total winding number becomes non-quantized even though the system is in the localized phase. Moreover, at this point $w$ changes its sign i.e. when $h <h_c$, $w$ is negative and when $h > h_c$, $w$ is positive. This analysis shows that the topological transition is also inline with the localization and real-complex transitions and shows a similar re-entrant behavior. Note that at $h_c$, a loop in the energy spectrum is expected since the system lies in localized phase at this point. However, as the largest eigenvalue is much larger compared to the other eigenvalues in the spectrum, a discontinuous loop is formed (not shown). This nature is also reflected in the value of $\langle r\rangle$ which decreases slightly from its maximum value of $\sim0.6$ (see Fig.~\ref{fig:fig1}(b)).

{\em Conclusions.-}
We have predicted a re-entrant topological transition in a non-Hermitian quasiperiodic AAH model in one dimension due to a re-entrant delocalization transition. We have shown that for the non-Hermitian AAH model with generalized quasiperiodic onsite potential, the system undergoes a delocalization-localizatio-delocalization transition as a function of the complex phase in the quasiperiodic potential. As the system undergoes the delocalization-localization-delocalization transition, the spectrum exhibits a real-complex-real transition indicating a $\mathcal{P}\mathcal{T}$ broken and subsequent unbroken phase transition. We have shown that as the localized phases exhibit complex spectra, a spectral winding number can be associated to the states making the localized phase topological. However, this winding number  vanishes in the delocalized phase as the entrire spectrum is real. Moreover, we have found that these transitions occur through two itermediate regions exhibiting both real and complex energy spectra. As a result, we have identified winding numbers in these interemediate regions which are different from the one obtained for localized region. 

The re-entrant topological transition in our paper is due to a non-trivial reappearance of the entire real spectrum as well as the delocalization of  the entire eigenstates in a non-Hermitian system where the non-Hermiticity is associated only with the onsite quasiperiodic potential and not on the hopping terms. This behaviour is in complete contrast to the existing predictions in other non-Hermitian quasiperiodic models where the system remains localized and the spectrum remains complex after the first localization transition. This prediction will enhance our understanding of the topological transitions in non-Hermitian translationally broken systems and also may lead to exploration of such re-entrant phenomenon in other interacting systems. On the other hand due to the recent progress in accessing quasiperiodic lattices in platforms such as photonic lattices and electrical circuits, our prediction can in principle be observed in experiments.

{\em Acknowledgment.-}
We acknowledge fruitful discussions with S.D. Mahanti, Awadhesh Narayan and M.J. Bhaseen. T.M. acknowledges support from Science and Engineering Research Board (SERB), Govt. of India, through project No. MTR/2022/000382 and STR/2022/000023. A.P. thanks NISER, Bhubaneswar for hospitality during which a part of this work has been done.

\bibliography{references}

\end{document}